\documentstyle[11pt,aaspp4]{article}

\lefthead{L\'epine et al.}  \righthead{Spatially resolved spectra of
WR+OB binaries}

\begin{document}

\title{Spatially resolved STIS spectra of WR+OB binaries with
colliding winds\altaffilmark{1}}

\author{S\'ebastien L\'epine\altaffilmark{2}, Debra
Wallace\altaffilmark{3}, Michael M. Shara\altaffilmark{2}, Anthony
F. J. Moffat\altaffilmark{4}, \& Virpi S. Niemela\altaffilmark{5,6}}

\altaffiltext{1}{Based on Observations with the NASA/ESA {\it Hubble
Space Telescope}, obtained at the Space Telescope Science Institute,
which is operated by the Association of Universities for Research in
Astronomy (AURA), Inc., Under NASA contract NAS 5-26555.}

\altaffiltext{2}{Department of Astrophysics, Division of Physical
Sciences, American Museum of Natural History, Central Park West at 79th
Street, New York, NY 10024, USA, lepine@amnh.org,mshara@amnh.org}

\altaffiltext{3}{Department of Physics and Astronomy, Georgia State
University, Atlanta, GA 30303, wallace@chara.gsu.edu}

\altaffiltext{4}{D\'epartement de Physique, Universit\'e de
Montr\'eal, C.P. 6128 Succ. Centre-Ville, Montr\'eal, QC, Canada, H3C
3J7, moffat@astro.umontreal.ca}

\altaffiltext{5}{Facultad de Ciencias Astron\'omicas y Geofisicas,
Universidad Nacional de La Plata, Paseo del Bosque s/n, 1900 La Plata,
Argentina, virpi@fcaglp.unlp.edu.ar}

\altaffiltext{6}{Member of Carrera del Investigador, CIC-BA, Argentina}

\begin{abstract} 
We present spatially resolved spectra of the visual WR+OB massive
binaries WR86, WR146, and WR147, obtained with the {\it Space
Telescope Imaging Spectrograph} on board the {\it Hubble Space
Telescope}. The systems are classified as follows: WR86 = WC7 + B0
III, WR146 = WC6 + O8 I-IIf, WR147 = WN8 + O5-7 I-II(f).
Both WR146 and WR147 are known to have strong non-thermal radio
emission arising in a wind-wind collision shock zone between the WR
and OB components. We find that the spectra of their O companions show
$H_{\alpha}$ profiles in emission, indicative of large mass-loss
rates, and consistent with the colliding-wind model. Our
spectra indicate that the B component in WR86 has a low mass-loss
rate, which possibly explains the fact that WR86, despite being a long
period WR+OB binary, was not found to be a strong non-thermal radio
emitter. Because of the small mass-loss rate of the B star component
in WR86, the wind collision region must be closer to the B star and
smaller in effective area, hence generating smaller amounts of
non-thermal radio emission. Absolute magnitudes for all the stars are
estimated based on the spectral types of the components (based on
the tables by Schmidt-Kaler for OB stars, and van der Hucht
for WR stars), and compared with actual, observed magnitude
differences. While the derived luminosities for the WC7 and B0 III
stars in WR86 are consistent with the observed magnitude difference,
we find a discrepancy of at least 1.5 magnitudes between the observed
luminosities of the components in each of WR146 and WR147 and the
absolute magnitudes expected from their spectral types. In both cases,
it looks as though either the WR components are about 2 magnitudes too
bright for their spectral types, or that the O components are about 2
magnitudes too faint. We discuss possible explanations for this
apparent discrepancy.
\end{abstract}


\section{Introduction}

OB stars (which as a group include stars of spectral type between O3
and B2) are the most massive main sequence objects. They are generally
found in young clusters and associations, because their lifespans are
too short to carry them far from their birthplaces. Consequently, they
are often subjected to considerable interstellar absorption.
Wolf-Rayet (WR) stars are believed to be the evolved counterparts of
at least some OB stars, and mostly correspond to a phase where the massive
star has lost all its external hydrogen envelope through stellar winds
(Conti 1976). Wolf-Rayet stars generate intense stellar winds reaching
rates of $\dot{M}\gtrsim10^{-5}$ M$_{\sun}$ yr$^{-1}$ and terminal speeds
$v_{\infty}\gtrsim10^3$ km s$^{-1}$. The Wolf-Rayet photosphere arises
not from the hydrostatic surface but in the wind itself which, besides
He lines, is dominated either by ions of Nitrogen (WN stars) or Carbon
and Oxygen (WC stars). Hence the spectrum of a WR star is dominated by
very broad lines of He and heavier elements.

Current estimates indicate that at least $\sim40\%$ of WR stars are
in multiple systems (Moffat 1995). The presence of a companion can be
suspected from a composite spectrum showing OB absorption lines or from
the apparent ``dilution'' of the WR emission lines by extra continuum
emission (Smith {\it et al.} 1996). Confirmation of multiplicity is
usually based on radial velocity studies. Indirect confirmation can
also be made by the detection of colliding wind effects (e.g. Tuthill
{\it et al.} 1999, Williams 1999). A small number of OB companions in
very long period ($P>50$ yr) systems have also been found from
speckle interferometry or direct imaging (Hartkopf {\it et al.} 1993,
Williams {\it et al.} 1997, Niemela {\it et al.} 1998). These systems
are particularly interesting because the absolute magnitude of the WR
component can be directly inferred from that of its companion,
assuming the two stars are at the same distance.

In the IR and radio regimes, WR stars are dominated by thermal,
free-free emission from the dense, expanding envelope. However,
$\approx40$\% of WR stars have been observed to have a radio
spectral energy distribution consistent with non-thermal emission
associated with synchrotron radiation and high-energy
phenomena. Eichler \& Usov (1993) have demonstrated how non-thermal
radio emission could arise from the collision between the outflows
from two early-type stars in a binary system. As it turns out, binary
systems are over-represented in the sample of non-thermal emitters,
which prompted van der Hucht (1992) to suggest that all non-thermal WR
emitters were actually in binary systems, with the wind collision
being responsible for the non-thermal emission.

The colliding wind model has been strikingly confirmed by radio
observations of two non-thermal WR emitters (Williams {\it et al.}
1997, Dougherty {\it et al.} 1996), confirmed to be in binary systems
with OB companions (Niemela {\it et al} 1998). The non-thermal
emission is found to be associated with a distinct region whose
photocenter is located {\em on or close to the line joining the WR
star and its OB companion}, the WR star itself being associated with a
thermal source. However, not all WR+OB binaries are found to be
non-thermal emitters. Dougherty \& Williams (2000) have noted that 
WR+OB systems form two distinct groups, with
most thermal emitters being in short period systems, while the
non-thermal emitters are all in long period systems, with large
component separation. Since the non-thermal emission arises locally,
at the wind-wind collision front, one might expect to observe
non-thermal radio emission only from widely separated WR+OB binaries,
whose large collision fronts are expected to be located well outside
the extended radio photosphere of the WR wind. In short period
binaries, the collision front occurs much deeper in the wind of the WR
star, in which case non-thermal radio emission is not expected to be
observed as the collision front lies inside the WR radio photosphere.
 
More recently, the conjecture that all non-thermal emitters are
colliding wind binaries has been put in doubt by Wallace {\it et al.}
(2000), who have failed to identify companions for 8 WR stars with
non-thermal radio emission, down to a mean projected separation of
$\approx$20 AU. Since this scale is within the distance of the radio
photosphere, either non-thermal radio emission in these stars arises
farther out in the wind itself (possibly generated by intra-wind
shocks), or non-thermal radio emission does arise in colliding winds
but the radio photosphere is much smaller than predicted. If the
latter hypothesis is correct though, all the apparently single
non-thermal radio emitters should have companions on close orbits, and
should have been identified as spectroscopic binaries. The single-star
explanation remains to be confirmed.

The colliding wind model, however, is successful in explaining
non-thermal emission in long-period WR+OB systems. But are all
long-period systems non-thermal emitters? So far, there is
one exception to the rule: the long period WR+OB system
WR86. However, while WR86 is not strictly defined as a non-thermal
emitter, its spectral index is consistent with a ``composite''
thermal/non-thermal source, i.e. it is consistent with weak
non-thermal emission (Dougherty \& Williams 2000).

This paper presents spatially resolved STIS spectra of the components in
the visual WR+OB systems WR86, WR146, and WR147. The observations and
data reduction are discussed in Section 2. Spectral classification of
the components is presented in Section 3. This allows us to estimate
both the absolute luminosity of the components, and the mass-loss
rates of the OB stars. Results are discussed in the light of the
colliding wind model in Section 4. We briefly summarize our findings
in Section 5.

\section{Observations}

Long-slit spectroscopy of the stars WR86, WR146, and WR147 have been
obtained with the STIS spectrograph on board the {\em Hubble Space
Telescope}. In each case, observations were carried out with the slit
length oriented as close as possible to the apparent position angle of
the binaries. Niemela {\it et al.} (1998) have estimated the position
angles to be PA=109$^{\circ}\pm$9$^{\circ}$, 21$^{\circ}\pm$4$^{\circ}$,
and 350$^{\circ}\pm$2$^{\circ}$ for WR86, WR146, and WR147,
respectively. The exact orientation of the STIS slit depends on the
orientation of HST at the time of the observation; hence the above PAs
were used as scheduling constraints. Observations were finally carried
out with PA=106.1$^{\circ}$,18.8$^{\circ}$, and 360.2$^{\circ}$, for
WR86, WR146, and WR147, respectively. In each case, the orientation
was such that the brightest component in the V band appeared on the
larger CCD column number. The width of the 52$\times$0.5 arcsec slit
is on the order of, or larger than, the separation between the
components; hence the small difference ($<13^{\circ}$) between the
orientation angle at the time of observation and the PAs of  the
systems has negligible effects on the throughput.

The two stars in the WR147 system are clearly resolved by STIS; we
used standard IRAF aperture extraction to obtain their spectra. On the
other hand, the relatively broad wings in the STIS point spread
function (PSF) resulted in the WR86 and WR146 spectra
being resolved, but with significant blending (Figure 1). To
complicate matters, the shape of the PSF was found to be dependent on
the wavelength, to a level that significantly affects the spectral
extraction of barely resolved sources like those from WR86 and WR146
(but is however of little consequence for clearly resolved sources
such as in WR147). We performed a multidimensional fit for each column
on the CCD, to extract a blended double profile $G(y)$ with the
general shape \footnote{The profile described by equation 1 is not a
gaussian, although is does assumes the gaussian form for the special
case $k=2$. We used this form for its simplicity and stability under
the multi-dimensionnal fit. We did attempt to use other simple
mathematical forms to describe the PSF, e.g. Moffat (1969) functions,
only to obtain equivalent results in the spectral extraction.}:
\begin{equation}
G(y) = A1 \exp{-(\frac{(y-y1)^2}{2 \sigma^2})^{k}} + A2
\exp{-(\frac{(y-y2)^2}{2 \sigma^2})^{k}} \ ,
\end{equation}
where $y1$ and $y2$ are the spatial locations of the stars along a
column, and $A1$ and $A2$ are the amplitudes of the profiles for each
star. The shape of the stellar profiles is governed by the dispersion
parameter $\sigma$ and the ``peakyness'' parameter $k$ (which sets the
relative strength of the PSF wings). Because the mean dispersion $\sigma$
of the PSF is found to be close to the pixel size, we re-sampled the
profile over individual pixels $y_i$ from the continuous function
$G(y)$ using:
\begin{equation} 
G'(y_i) = \int_{y_i-0.5}^{y_i+0.5} G(y) dy \ ,
\end{equation}
where $y_i$ are integers representing the CCD lines.

We then used the assumption that the separation between the stars
should be constant and independent of the wavelength. To determine the
separation between the stars, we first performed a fit of $G'(y_i)$ in
the 6-dimensional parameter space $(y1,y2,\sigma,k,A1,A2)$ and
calculated the separation $\Delta y$ between the stars from the mean
values of $y2-y1$ obtained from each CCD column. We then fixed
$y2=y1+\Delta y$ and performed a second (more robust) fit in the
5-dimensional space $(y1,\sigma,k,A1,A2)$, for each CCD column.

Using the derived values for $\Delta y$ and the STIS pixel size of
0.0507 arcsec pixel$^{-1}$, we estimate the separation along the slit
between the components of WR86, WR146 and WR147 to be
0.239$\pm$0.006$^{\arcsec}$, 0.161$\pm$0.005$^{\arcsec}$, and
0.624$\pm$0.015$^{\arcsec}$, respectively. The PA of the slit was
always within 13$^{\circ}$ of the line joining the two stars, hence
these values should be reliable estimates of the component
separations. Our values for WR146 and WR147 are consistent with the
separation calculated from the WFPC2 image by Niemela {\it et al.}
(1998). For WR86, our derived separation is outside of 1$\sigma$
of the 0.286$\pm$0.039$^{\arcsec}$ range derived by Niemela {\it et
al.}, but well within the 0.230$\pm$0.013$^{\arcsec}$ range cited
in the HIPPARCOS catalog.

Results of the multidimensional fits confirm the existence of a
systematic variation in the PSF pattern with wavelength (Figure
2). The pattern also differs between WR86 and WR146, despite the same
spectral coverage, which suggests that the shape of the PSF
also depends on where the source falls along the slit (the WR86 and
WR146 systems have been recorded on slightly different CCD
columns). This raises the possibility that the PSF may be slightly
different for each component in any one system, which may result in
some inaccuracies in the spectral extraction.

The two stellar components in both WR86 and WR146 were apparently
separated reasonably well across most of the spectral range. The only
exception occurs at wavelengths close to the very bright WR emission
line CIV $\lambda$5808 (at which point the WR star is significantly
brighter than the OB star), where a small contamination of the WR flux
onto the OB companion spectrum apparently occurred. This is clearly
due to an imperfect PSF model. In particular, the wings of the PSF
showed evidence for a weak diffraction pattern, which our model does
not reproduce. This results in some contamination of one spectrum by
some light from the other component. However, this contamination is
apparent only for the broad CIV emission line, because of the large
difference in the brightness of the WR and OB components at that
wavelength, and is negligible elsewhere (below instrumental noise
levels). Fortunately, this contaminated CIV emission from the WR
component can be unambiguously identified in the OB star spectra,
whose spectrum does not normally exhibit such a broad emission feature
at that wavelength. Because the spectra are generally well separated,
and because contamination effects are relatively small, we did not
attempt to refine the PSF model further, which would have required the
use of extra parameters and would have made the multi-dimensional fit
much more difficult to perform.

The resolved spectra for all three stars are shown in Figure 3. It
turns out that the brightest component in the V band in each pair is
the Wolf-Rayet star, even though the continuum emission from the OB
star is actually larger in WR86 and WR146 (the WR stars in these
systems are brighter in V because of their strong emission
lines). Contamination of the OB spectra by the very bright CIV
$\lambda$5812 line is very obvious in WR146 (see Figure
5). Examination of the CIV $\lambda$5812 contamination profile in the
spectrum of the O component shows that contamination becomes
apparent as the monochromatic intensity from the WR star reaches
$\approx3$ times that of the O companion. Since the relative WR
intensity is below that level over the remainder of the spectrum, we
conclude that contamination effects must be negligible over the
remainder of the spectral range. Hence, any feature observed elsewhere
in the spectrum of the O star is assumed to be intrinsic. In WR86, the
components are further apart; hence the extracted spectra are less
susceptible to contamination effects. Our extracted spectra of
WR86 show evidence for a weak contamination of the CIV$\lambda$5812
line, and also possibly from CIII $\lambda$5696 (see Figure 7). Both
WR lines have monochromatic intensity reaching $\approx$3 times that of
the B star continuum. Because these lines are the brightest
features in the WR component of WR86, and because they yield only weak
contamination effects, we conclude that no other WR features
contaminates the B spectrum significantly, and any other feature
observed in the B star spectrum must be intrinsic. 

In each system, the slopes of the continua from the WR stars and the
OB companions as well as the strength of the interstellar absorption
lines are all consistent with equal amounts of reddening. This
supports the idea that the components in each system are approximately
at the same distance, and most likely to be physically related. In
both WR146 and WR147, we confirm that the WR (O) component is to the
south (north), consistent with the colliding wind interpretation of
their radio maps (see Niemela {\it et al.} 1998). For WR86, the WR
component is to the north-west and the B component to the south-east.

We did not find any trace of background, or ``nebular'' emission in
the longslit spectral images, within the instrumental limits. Attempts
have been made to extract spectra at different locations along the
slit, but only the wings of the PSF from the stellar components and
other instrumental features were detected. This lack of detection is
significant for WR146 and WR147, which are known to be colliding wind
binaries. If there is any diffuse emission in the optical associated
with the colliding wind region, we estimate that it must be weaker
than 5 10$^{-15}$ ergs cm$^{-2}$ s$^{-1}$ \AA$^{-1}$ arcsec$^{-1}$.

Line identifications for each of the WR and OB star components are
listed in Tables 1-6. The resolution of the STIS spectra was
$\approx1.4$\AA, which is the accuracy in the central wavelength
measurements. Estimated errors on equivalent width measurements
$W_{\lambda}$ are listed individually. The error on $W_{\lambda}$
depends on the strength of the line, and on whether it was resolved or
in a blend. We do not give the mean central wavelengths of the lines
in the WR stars, because the very broad profiles make the central
wavelengths very unreliable for line identification. Most of the
bright WR lines are actually blends of several different lines; the
line identification and rest wavelength given in the tables is for
the transition which most probably contributes to the largest part of
the line emission.

\section{Spectral Classification}

\subsection{WR86}

This is the V=9th magnitude system HD156327, located at $\alpha=$17 18
23.06 $\delta=$-34 24 30.6 (J2000). It was initially listed as a
Wolf-Rayet binary with spectral type WC7+B0 V (Roberts 1962; Smith 1968),
based on the presence of H and HeI absorption lines in the blue. It
was included in the Sixth Catalog of Galactic Wolf-Rayet stars (van
der Hucht {\it et al.} 1981) under the designation WR86, and given a
WC7+abs spectral type, implying that absorption lines could be intrinsic to
the WR star, thus questioning its double star status. However, HD156327
was known to be a close visual double with separation
$\rho\sim0.2\arcsec$ (Jeffers {\it et al.} 1963). The fact that Massey
{\it et al.} (1981) failed to measure any radial velocity variation
ruled out the idea of a {\it close} OB companion, strongly suggesting
that the OB spectrum is associated with the visual companion (see
Moffat {\it et al.} 1986). In any case, the star continued to be
referred to as a WC7+abs throughout the 1980s.

The double star status was confirmed with speckle observations by
Hartkopf {\it et al.} (1993), who resolved the star into two
components with a $0.237\arcsec$ separation. The components were also
clearly resolved by the WFPC2 camera on board HST (Niemela {\it et
al.} 1998). Based on scanned image tube spectra of the pair,
Niemela {\it et al.}  suggested the companion to be a B0 star
(detection of OII, SiIII, and SiIV) with a luminosity class between I
and III. The system is now listed in the seventh catalog of Wolf-Rayet
stars (van der Hucht 2001) as ``WC7 (+B0 III-I)''.

Our STIS spectra confirm that the WR component is of subtype WC7
(Figure 4, with line equivalent widths listed in Table 1). The ratio
of the equivalent widths of the CIV $\lambda$5801 and CIII
$\lambda$5696 lines is log W$_{\lambda}$(CIV 5801)/log
W$_{\lambda}$(CIII 5696) $\simeq$ 0.16, which is consistent with
subtype WC7 in the quantitative classification system of Crowther {\it
et al.} (1998).

For the O component (Figure 5; Table 2), our spectra confirm the B0
type, based on a comparison of the blue spectrum with the atlas of
Walborn \& Kirkpatrick (1990). We attempt to better constrain the
luminosity class based on the strength of the H$_{\gamma}$ absorption
line, for which a calibration with the absolute magnitude has been
derived by Millward \& Walker (1985). We measure
$W(H_{\gamma})=2.60\pm0.15$\AA\, the uncertainty being largely
attributable to the blend with OII $\lambda$4349. Following the
Millward \& Walker calibration, this corresponds to an absolute
magnitude $M_V\simeq-4.8\pm0.2$. According to the B-star absolute
magnitude calibration of Schmidt-Kaler (1982), this makes the star a
B0 giant of class III. Since the WFPC2 photometry shows the two stars
to have the same $M_V$ (within the $\pm0.09$ observational errors),
then the WR component is also estimated to have
$M_V\simeq-4.8\pm0.2$. This value is largely consistant with the range
of values quoted by van der Hucht (2001) for WC7 stars.

Luminosity classes of early B stars can also be estimated from the
ratio of SiIII $\lambda$5740 and HeI $\lambda$5876. Comparison with the
yellow-red atlas of Walborn (1980) shows the spectrum to be
generally consistent with luminosity class III. While we can
definitely rule out a class I for this object, it is not possible to
rule out spectral class II on the basis of the SiIII $\lambda$5740 /
HeI $\lambda$5876 ratio. However, because the $H_{\alpha}$ line
shows no trace of overlying wind emission (which occurs in most early
B stars with luminosity class I-II), we classify this system as WC7 + B0 
III.

\subsection{WR146}

This star, located at $\alpha=$20 35 47.09 $\delta=$+41 22 44.7
(J2000), was initially classified as WC6 by Roberts (1962), and as WC5
by Smith (1968). It was listed as a WC4 in the sixth catalog of
Galactic Wolf-Rayet stars (van der Hucht {\it et al.} 1981). Improved
measurements of the line ratios led Smith {\it et al.} (1990) to
reclassify the WR star as WC6.

Dougherty {\it et al.} (1996) observed the star at 1.6 GHz and 2.5 GHz
with the MERLIN array, and resolved the source into two components, a
thermal source and a non-thermal source, separated by $116\pm14$
milliarcseconds. They attributed the thermal source to the WR star,
and the non-thermal source to a colliding-wind region between the WR
star and an OB companion. Dougherty {\it et al.} (1996) also found
weak hydrogen absorption lines ($H_{\delta}$, $H_{\gamma}$) in an
unresolved blue spectrum of WR146, which they attributed to the
unresolved companion.

Optical WFPC2 images from HST clearly resolved the object into two
components separated by $168\pm31$ arcsecs, and with very similar
colors (Niemela {\it et al.} 1998). An overlap of the optical images
and radio maps showed the non-thermal source to be located between the
optical components, confirming the colliding-wind binary
hypothesis. Assuming that the relative location of the non-thermal
source matches the head of the bow shock, it is possible to calculate
the ratio of the wind momentum fluxes. For WR146, Niemela {\it et al.}
obtained a ratio $\eta \equiv
(\dot{M}v_{\infty})_{OB}/(\dot{M}v_{\infty})_{WR}\sim0.1$.  Given the
large mass-loss rate expected from the WC6 star, this requires the
companion to have a relatively large mass-loss rate also. Based on the
momentum ratio and on the relative colors of the components, Niemela
{\it et al.} suggested the companion to be O6-O5 V-III.

More recently, Dougherty {\it et al.} (2000) obtained a $3800-4500$\AA\
spectrum of WR146 at the 4-m William Herschel Telescope (WHT). Though
blended with emission lines from the WR star, the relatively narrow
absorption lines from the O companion were clearly identified. Their
$\lambda$4541 HeII / $\lambda$4471 HeI equivalent width ratio (the
principal diagnostic of the O-type sequence) indicated a spectral type
O8. Though several features were also suggestive of a high luminosity
class, they did not assign a luminosity class due to a lack of the
main luminosity diagnostic lines in their spectrum. The system is now
listed as WC6+O8 in the seventh catalog of Wolf-Rayet stars (van der
Hucht, 2001).

Our STIS spectra of the WR component (Figure 6; Table 3) confirms the
WC6 classification. We measure a line equivalent width ratio log
W$_{\lambda}$(CIV 5801)/log W$_{\lambda}$(CIII 5696) $\simeq$ 1.03,
consistent with subtype WC6 in the quantitative classification system
of Crowther {\it et al.} (1998). The lines in this star are especially
broad, indicating a large wind terminal velocity
($v_{\infty}\simeq2900$ km s$^{-1}$ as measured by Eenens \& Williams
1994).  

For the OB component (Figure 7; Table 4) we measure a line
ratio He II $\lambda$4541 / He I $\lambda$4471 = 0.39 which is
consistent with a spectral type O9 in the system of Conti \& Alschuler
(1971) and Conti (1973). The HeI $\lambda$4471 line does look
significantly broader than HeII, which may be due to noise, but could
also result from blending of the HeI $\lambda$4471 with some other
unidentified line. On close examination, the HeI $\lambda$4471
profile in the B0 III component of WR86 does look asymmetric, with the
blue side of the line being unusually extended. If this is due to some
unidentified IS absorption feature (we do indeed see a shallow
absorption trough at the same wavelength in the WR spectrum which
could also be the signature of this IS feature, see Figure 6), then
the EW of this line is most likely to be overestimated in WR146. We
consider the HeI $\lambda$4388 line, and note that it is clearly
weaker than HeII  $\lambda$4540; the spectral atlas of Walborn \&
Fitzpatrick (1990) shows that this generally does not occur in O9 type
stars, where both lines have about the same EW. It is however
consistent with spectral type O8, and we thus also adopt this
classification for the O star in WR146.

The main luminosity diagnostic for O8-O9 stars is the increase in the
strength of the Si IV $\lambda\lambda$4089, 4116 at higher
luminosities, and the change in NIII $\lambda\lambda$4634, 4640 and He
II $\lambda$4686 from absorption to emission (Walborn \& Fitzpatrick,
1990). While the WHT spectrum of Dougherty {\it et al.} included both
Si IV lines, they were found to be blended with the $H_{\delta}$
line. Unfortunately, neither the STIS nor the WHT spectrum covers the
He II $\lambda$4686 region. However, we note the presence of a weak
emission feature centered on $\lambda$5696 which we attribute to CIII
line emission. Both the CIII $\lambda$5696 and H$_{\alpha}$ lines have
been shown by Walborn (1980) to behave like NIII $\lambda\lambda$
4634, 4640 and He II $\lambda$4686, respectively.

The H$_{\alpha}$ profile in the O star component shows a narrow but
relatively shallow absorption trough flanked by relatively broad
wings. The H$_{\alpha}$ line in the O star is coincident with the HeII
$\lambda6560$ complex in the Wolf-Rayet star, which raises the
possibility of contamination into the O spectrum, which would explain
the broad wings. However, we argue that the broad H$_{\alpha}$
emission profile is intrinsic to the O star: (1) the intensity of the
HeII $\lambda6560$ complex relative to the O star continuum is
comparable to that of the broad shoulder in the CIV $\lambda5812$
complex (between $\lambda5850$ and $\lambda5950$), and we see no
obvious contamination of the O star spectrum by the latter, (2) the
H$_{\alpha}$ profile in the B0 component in WR86 does not show any
broad wings, while it was more prone to being contaminated by the
more intense HeII $\lambda6560$ complex from its WR component, (3) the
centroid of the H$_{\alpha}$ absorption trough is 6.4\AA larger than
the measured lab value, which suggests that the central trough is
being distorted as the line is filled up with H$_{\alpha}$ in
emission, and (4) the overall shape of the H$_{\alpha}$ profile is
very similar to that of other OIf stars (e.g. Cygnus OB\#7 and
HD210839 see Figure 6 in Herrero {\it et al.} 2000), including both
the central narrow absorption and the broad emission wings.

The behavior of CIII $\lambda$5696 and H$_{\alpha}$ in the O component
of WR146 clearly rules out any luminosity class fainter than
II. Because the H$_{\alpha}$ line is not found to be strongly in
emission, as in O8 I stars, a spectral type O8 II is the most
reasonable classification. A secondary indicator for the luminosity
class is the ratio He I $\lambda$4388 / He I $\lambda$4471, which
correlates with the mass-loss rate. Our spectrum yields He I
$\lambda$4388 / He I $\lambda$4471 = 0.2, which suggests that the
star is Of. The weakness of H$_{\gamma}$ ($W_{\lambda}=1.5$) is
also consistent with large intrinsic luminosity; the Millward \&
Walker (1985) relationship suggests $M_{V}\gtrsim-6$ which, for O8, is
consistent with luminosity class I-II (Schmidt-Kaler 1982). From these
lines of evidence, we classify this system as WC6 + O8 I-IIf.

\subsection{WR147}

This star, located at $\alpha=$20 36 43.65 $\delta=$+40 21
07.3 (J2000), was resolved into a double radio source with a thermal lobe
and a non-thermal lobe by Churchwell {\it et al.} (1992). Williams
(1996) hypothesized that the non-thermal lobe was the result of a
colliding-wind interaction with an unseen companion. The companion was
first identified as a faint component in an IR image of WR147 
(Williams {\it et al.} 1997), and confirmed in the optical by HST
WFPC2 observations (Niemela {\it et al.} 1998). Based on the relative
K magnitude between the WN8 star and the IR companion
($K_{OB}-K_{WR}$=3.04$\pm$0.09), and assuming the WN8 star to have an
absolute magnitude M$_{K}\approx-6$, Williams {\it et al.}  claimed
the OB companion to be consistent with a spectral type B0.5 V. In the
optical, the magnitude difference between the two components is
observed to be $V_{OB}-V_{WR}$=2.16$\pm$0.09, which is too small for a
B0.5 V companion, and Niemela {\it et al.} (1998) proposed that the
star is of earlier type.

Our STIS data for the WR component in WR147 (Fig. 8; Table 5) shows
a spectrum typical of a WN8 star, with a relatively strong P Cygni
profile in HeI $\lambda$5876. Also present are the weak HeII
$\lambda\lambda$6311, 6406, and 6527 lines \---- all with P Cygni
profiles. Weak NII $\lambda\lambda$5680, 5686 lines are also detected,
consistent with a late-type WN star.

Our WR147 spectrum in the blue regime is of very poor quality and is
not shown in this paper. Because of the considerable interstellar
reddening, the WR147 system is several magnitudes fainter in the blue,
and our HST STIS exposures were not programmed appropriately for this
relatively faint target. Hence it is not possible to obtain a precise,
reliable spectral type for the O component since classification of OB
stars is largely based of the ratio of HeI and HeII lines in the
blue. However, our spectrum shows a very weak HeI $\lambda$5876 line,
while HeI $\lambda$6678 is too weak to even be detected (Fig.9; Table
6). Since HeI is very strong in late O stars and early B stars (type
B2 being where HeI reaches a maximum; see Walborn \& Fitzpatrick
1990), this suggests that this star is either earlier than O8, or
mid/late B. Because He becomes doubly ionized in early O stars, we can
also exclude a spectral type earlier than O5. Furthermore, the
presence of high ionisation CIV lines unambiguously rules out any
spectral type B or later. These lines of evidence suggests that our
star can only be in the range O5-O7. An O5-O7 spectral type is
consistent with the relatively strong CIV $\lambda\lambda$5801,5812
lines, which are strongest in this spectral range (Walborn
1980). Because we lack a clear line ratio, and because there is
considerable scatter among O stars in the equivalent widths of single
species, it is however not possible to constrain the spectral type
further. Due to the lack of other objective classification criteria,
the O5-O7 assignment must be regarded as tentative.
 
The H$_{\alpha}$ region shows a very shallow absorption trough,
flanked by broad emission wings, very similar to the profile observed
in the O8 component in WR146. The broad emission wings cannot
be explained by contamination from the WN8 star, since the two
components are well resolved by STIS and any (very weak) component
arising in the very broad wing of the STIS PSF from the WN8 star should
be subtracted out with the background in the aperture extraction
procedure. Furthermore, the centroid of the observed absorption trough
is 8.9\AA over the H$_{\alpha}$ laboratory wavelength; this suggests
that the line profile is significantly distorted as it is filled up
with H$_{\alpha}$ emission arising in the wind. Finally, the overall
shape of the H$_{\alpha}$ profile is very similar to that of other O
If stars (e.g. Cygnus OB\#7 and HD210839, Herrero {\it et al.} 2000).

This $H_{\alpha}$ emission suggests a substantial mass loss
rate, which would imply that the star is a supergiant in the Of
category. However, CIII $\lambda$5696 is not clearly in emission; this
is more consistent with stars in the (f) category, which show a
filling in of HeII $\lambda$4686 but no strong emission lines. It is
therefore not possible at this point to clearly distinguish between
spectral class Ia, Iab, Ib, or II, but luminosity classes III-V can be
excluded. We tentatively classify this star as O5-7 I-II(f).

\section{Discussion}

\subsection{On the absolute magnitudes of the components}

Absolute magnitudes $M_{\it v}$ (in the narrowband photometric system
defined by Westerlund 1966) have been estimated for Galactic WR stars
on the basis of cluster and association membership (van der Hucht
2001, hereafter vdH01). The narrowband $v$ is used because it avoids
the brightest optical emission lines of WR stars. The distances of
clusters and associations derived by L\"undstrom \& Stenholm (1984)
are used. The mean absolute magnitudes $M_v$ for WC7, WC6, and WN8
stars are $M_v\simeq$ -4.5, -3.5, and -5.5, respectively, with a
standard deviation $\lesssim1.0$ magnitude. 

The absolute magnitudes $M_V$ of the OB stars can be estimated, based
on their spectral classification, using the relationships defined by
Schmidt-Kaler (1982, hereafter SK82). While the $v$ and $V$ bands are
not exactly the same, a comparison between $V$ and $v$ for a number of
WR and OB stars shows that $|v-V|\lesssim0.3$ (Westerlund 1966, see
also Turner 1982). While the value of $v-V$ for any given WR star will
be dependent on the strength of the optical lines, we can assume that
$M_v \sim M_V$ to within less than half a magnitude.

We thus compare the difference in absolute magnitudes as determined
from the spectral types with the measured difference in $V$ derived
using the WFPC2 images of these systems by Niemela {\it et al.}
(1988). For the WR86 system, the WC7 star is expected be in the range
$-5.5<M_V<-3.5$, and SK82 quotes a value of $M_V\simeq-5.1$ for a B0 III
star. The two estimates are thus consistent with each other, since the
observed difference in $V$ magnitude for this system is
$V_{WR}-V_{B}=0.02\pm0.18$.

For the WR146 system, the WC6 star is expected to be in the range
$-4.5<M_V<-2.5$, while SK82 quotes a mean value of $M_V\simeq-6.25$
for O8 I-II stars. The observed magnitude difference is $V_{WR}-V_{B} =
-0.24\pm0.08$; there is thus a discrepancy of at least 2 magnitudes
between the observed magnitude difference and that inferred from
the spectral types. The two stars are definitely companions, as
evidenced by the confirmation of a colliding wind region between the
two components. The discrepancy thus cannot be attributed to a chance
alignment of two stars at different distances. If we assume that the
O8 star is actually a main sequence object (unlikely because of the
$H_{\alpha}$ line in emission), we get $M_V=-5.15$ from SK82, which
still yields a difference of at least one magnitude. This means either
that the WR star is too bright for its spectral type or, alternately,
that the O star is too faint. An intriguing comparison can be made
between WR146 and WR86: both systems have components of approximately
the same magnitude, but the WC5 star in WR146 is expected to be
intrinsically fainter than the WC7 star in WR86, while the O8 I-II
star in WR146 is expected to be intrinsically brighter than the B0 III
star in WR86. 

For the WR147 system, the WN8 star is expected to be in the range
$-6.5<M_V<-4.5$. From SK82, we get a range of values $-5.90<M_V<-6.25$
for O5-7 stars with luminosity classes I-III. The observed magnitude
difference is $V_{WR}-V_{B} = -2.16\pm0.12$. Again, we find that the
WR component is too bright for its spectral type by at least 1.5
magnitudes or, alternately, that the O star is too faint by at least
1.5 magnitude. To be consistent with the expected magnitude of a WN8
star, the O star would have to be fainter than $M_V\approx-4.4$ which,
in the SK82 tables, is only consistent with stars of type B0 or
later. A spectral type of B is clearly ruled out by our STIS spectra.

Recently, absolute magnitudes of OB stars were re-evaluated, based on
data by HIPPARCOS (Wegner 2000, hereafter W00). The values of $M_V$
for giant and supergiant OB stars derived by W00 turn out to be 
fainter than the SK82 values by about 2 magnitudes. Under the W00
system, the absolute magnitudes of WR146 and WR147, as estimated from
the spectral types would be consistent with the observed magnitudes.
Although the agreement is suggestive, one has to be concerned about
the possible effects of the revised $M_V$ from the W00 system on the
distance to Galactic clusters and associations, and hence on the
absolute magnitudes derived by vdH01. One also has to consider the
relatively large scatter in the absolute magnitudes of individual OB
stars as derived from HIPPARCOS parallaxes. For O7-B0 stars, the
scatter can be as large as $\pm1$ magnitude. There may also be
systematic effects on parallax estimates of the OB star HIPPARCOS
sample, especially since most of the OB stars are beyond 200pc. Hence,
the values derived from W00 must be used with caution.

Some of the discrepancy may arise because of the uncertainty in the
determination of the spectral-type of the OB components. In
particular, the main criterion used in the determination of the
luminosity class was the apparent filling in of the $H_{\alpha}$ line,
which suggests high mass-loss rates in WR146 and WR147. However, both
WR146 and WR147 are systems with colliding winds. If $H_{\alpha}$
emission arises in the shock front, this could fill-up the
$H_{\alpha}$ absorption line in the spectra of the OB companions,
mimicking the effect of intense mass-loss. If $H_{\alpha}$ emission
arises near the head of the bow shock front, which is relatively close
to the O star in both WR146 and WR147, then we would not be able to
resolve them from the O star in the STIS data presented here, and this
might explain the apparent filling in of the emission lines. On the
other hand, if $H_{\alpha}$ emission was to arise downwind from the
head of the bow shock, at some relatively large distance from both the
WR and O components, then any extra $H_{\alpha}$ emission would appear
as a diffuse component on the STIS data, which would not be strong
enough (as we have shown from the lack of any detected diffuse
emission, see section 2) to account for the apparent filling in of the
emission lines.

\subsection{On the binary status and the wind collision}

The WR146 and WR147 systems exhibit strong non-thermal radio
components. It has been found that the non-thermal emission occurs
{\em between} the two stellar components, somewhat closer to the OB
star (Dougherty {\it et al.} 1996, Williams {\it et al.} 1997). In the
framework of the colliding wind model developed by Eichler \& Usov
(1993), the shock front forms at a distance $r_{OB}$ ($r_{WR}$) from
the OB (WR) component. Given $D=r_{OB}+r_{WR}$ the distance between
the two stars, then: 
\begin{equation}
r_{OB}=\frac{\eta^{1/2}}{1+\eta^{1/2}} D ,
\end{equation}
where $\eta \equiv (\dot{M}v_{\infty})_{OB}/(\dot{M}v_{\infty})_{WR}$
is the wind momentum ratio. The ratio $r_{OB}/D$ is dimensionless and
independent of the projection of the system on the plane of the sky,
and thus can measured directly from imaging. 

For WR146 and WR147, $r_{OB}/D \simeq$ 0.25 and 0.14, respectively, as
measured directly from the combined radio maps and HST WFPC2 images
(Niemela {\it et al.} 1998). These yield estimated values of
$\eta=0.10$ and $\eta=0.028$ for WR146 and WR147, respectively. From
the spectral types, we estimate that the O components in WR146 and
WR147 have mass-loss rates $\dot{M}_{O}\sim1 10^{-5}$ M$_{\sun}$
yr$^{-1}$ (see Herrero {\it et al.} 2000). If we assume a typical
terminal velocity $v_{\infty}\approx$2000 km s$^{-1}$ for the O stars
(see Prinja, Barlow \& Howarth 1990), we can estimate the mass-loss
rate of the WR components and check for consistency. For WR146, taking
$v_{\infty}=$2900 km s$^{-1}$ as the terminal velocity of the WC6
star, this yields a mass-loss rate $\dot{M}_{WR}\sim$6.9 10$^{-5}$
M$_{\sun}$ yr$^{-1}$. This value is consistent with the mass loss rate
value of $\dot{M}_{WR}=3.0\pm1.5 10^{-5}$ derived from the radio
emission of the WR star by Dougherty {\it et al.} (1996). For WR147,
however, taking $v_{\infty}=$1100 km s$^{-1}$ as the terminal velocity
of the WN8 star (from Eenens \& Williams 1994), one gets $\dot{M}_{WR}
\sim $6.5 10$^{-4}$ M$_{\sun}$ yr$^{-1}$, which is an order of
magnitude larger than the mass-loss rate value of $\dot{M}_{WR}=4.6
10^{-5}$ estimated from the radio emission by Williams {\it et al.}
(1997).

For WR147, Williams {\it et al.} (1997) have deduced a distance of
$\approx630$pc by comparing infrared photometry of WR147 and WR105,
the latter being another WN8 star suspected to be in the Sgr OB1
association, whose distance is known. It is from this estimated
distance and from VLA measurements of the radio flux that Williams
{\it et al.} (1997) have estimated the mass-loss rate in the WR component
of WR147 to be 4.2$\pm$0.2 10$^{-5}$ M$_{\sun}$ yr$^{-1}$. The
disagreement with our estimated value can be resolved in two ways:
(1) we have overestimated $\dot{M}_{O}$ by at least an order of
magnitude, in which case the OB component in WR147 is unlikely to be an
O supergiant, or (2) WR147 is more distant than estimated by Williams
{\it et al.} (1997), i.e. at least 1 kpc away. If the latter
interpretation is true, then there exist significant differences between
the spectral energy distributions of WR105 and WR147. The former
interpretation is more likely to be true, which means that we have
overestimated the mass-loss rate of the O component because of an
inaccurate assessment of the spectral type and luminosity class. The
key element here is the observation of the filling in of the $H_{\alpha}$
profile, which does suggest a high mass-loss rate and
luminosity. Clearly, a resolved spectrum of the OB component in WR147
spanning the whole optical range is required to resolve this
issue. Such an observation will have to be performed with the Hubble
Space Telescope, or with similar high spatial resolution from the
ground.

While WR86 does show evidence for some non-thermal emission, it is
listed by Dougherty \& Williams (2000) as having a ``composite''
spectral energy distribution (as compared to a ``non-thermal''
spectral energy distribution for WR146 and WR147, and a
``thermal'' energy distribution for suspected single WR stars). We
believe that the fact that this star has only a weak component of
non-thermal emission can be explained by the smaller mass-loss rate from
the B component. While late Of stars typically have derived
mass-loss rates $\dot{M} \sim 10^{-5}$ M$_{\sun}$ yr$^{-1}$, a star of
spectral type B0 III like the companion to WR86 is expected to have a
mass-loss rate $\sim 10^{-7}$ M$_{\sun}$ yr$^{-1}$, or about two orders of
magnitude smaller than Of stars (e.g. Runacres \& Blomme
1996). Assuming $V_{\infty}$ for the B0 III star to be of the same
order of magnitude as for the supergiant O stars in WR146 and WR147,
this means that the wind momentum from the B0 III star must be two orders
of magnitude smaller and $r_{OB}$, the distance from the OB star to
the wind collision front, is expected to be one order of magnitude
smaller for the WR86 system than it is for WR146 and WR147. This means
that the WR86 shock front is formed much closer to the OB component
and most likely wraps around the OB component with a much smaller
opening angle. Hence, the total volume where non-thermal emission
occurs should be significantly smaller, which accounts for the weaker
non-thermal emission. This hypothesis can be tested by imaging the
WR86 system in the IR/radio at very high spatial resolution, and
locating the source of the weak non-thermal emission. We predict that
the non-thermal emission occurs very close (a few milliarcseconds) to
the OB component, where $r_{OB}/D\sim0.01$.

It has been suggested by Dougherty et al. (2000) that some apparent
discrepancies in the luminosities of the components in WR146 might be
resolved if the O8 companion was itself a WR+O binary. Likewise, Setia
Gunawan {\it et al.} (2000) have interpreted the observed 3.38 yr
period in the 1.4 GHz radio emission as evidence for a third component
in the system orbiting the O8 star. Our resolved spectra of WR146 shows
no clear evidence for a third component. It is true that the CIV
emission feature in the O8 star spectrum, which we marked as
contaminated light from the WC6 star, does look significantly different
from the actual CIV profile in the WC6 spectrum, and thus one may argue
that the so-called "contaminated light" could actually be the
signature of another WR star orbiting the O8 component. We note however
that (1) the fact that the PSF is strongly dependent on the wavelength
most likely introduces a dependency on wavelength for the amount of
contaminated light which can distort the CIV profile on the O8 star,
and (2) the extracted spectra being the result of a multidimensional
fit of the whole double PSF profile, we do not necessarily expect the
contamination to add up in a strictly linear fashion (i.e. effects may
be non-linear), which means that a disproportionately strong
contamination could occur at the point where the CIV line is
brightest, hence distorting the CIV profile in the O8 star spectrum.
We therefore conclude that there is no evidence for another WR star
orbiting the O8 star component on a tighter orbit. We can safely rule
out the presence of any unresolved WR star, except for one that would
be significantly fainter (at least 1 magnitude) than the resolved WC6
component. Unless the WC6 is unusually bright for its spectral type,
this leaves only a relatively faint WR star of spectral type WC3-WC4 or
WN2-WN3 (see van der Hucht 2001) as a possible (but unlikely)
candidate. If the O8 star is an unresolved double, it is more likely
to be an OB+OB system.

\section{Summary}

We briefly summarize our findings as follows:

\begin{enumerate}

\item We have obtained resolved spectra of the components in the close
visual binary systems WR86, WR146, and WR147. WR86 is classified as
WC7+B0 III, with the WR component to the northwest and the B component
to the southeast. WR146 is classified as WC6 + O8 I-IIf, with the WR
component to the south and the O component to the north. WR147 is
classified as WN8 + O5-7 I-II(f), with the WR component to the south and
the O component to the north. The relative location of the WR and O
components in the WR146 and WR147 systems is consistent with the
colliding wind interpretation of their radio maps.

\item Absolute magnitudes $M_V$ of the OB stars have been derived
based on the spectral type-magnitude relationship of Schmidt-Kaler
(1982), and compared with the estimated absolute magnitudes $M_v$ of
the Wolf-Rayet stars (from van der Hucht 2001). While the values are
consistent for the WR86 system, we find a significant discrepancy in
the WR146 and WR147 systems. For WR146, it looks like the WC6
star is at least 2 magnitudes brighter than expected (or the O8 I-IIf
star is at least 2 magnitudes fainter than expected). For WR147 the
WN8 star appears to be at least 1.5 magnitudes brighter than expected
(or the O5-7 I-II(f) star is fainter than expected).

\item From the spectral types, we estimate that the O components in
WR146 and WR147 have mass-loss rates $\dot{M}_{O}\sim10^{-5}$
M$_{\sun}$ yr$^{-1}$. These values can be compared to the estimated
values for the WR component mass-loss rate $\dot{M}_{WR}$, which are
linked to $\dot{M}_{O}$ through the configuration of the
colliding-wind systems. While the estimated value of $\dot{M}_{O}$ for
WR146 is consistent with $\dot{M}_{WR}$, our value of $\dot{M}_{O}$
for WR147 is an order of magnitude too large. This most likely
indicates that our spectral classification is inaccurate, although it
could also mean that current estimates of the distance to WR147 are
too low. A more accurate spectral classification for WR147 is required
to resolve the discrepancy, which will require new resolved
spectroscopic observations of the OB component in WR147.

\item From the spectral type, we estimate the B component in WR86 to
have $\dot{M}\sim10^{-7}$ M$_{\sun}$ yr$^{-1}$. The relatively smaller
mass-loss rate in the OB component in WR86 must result in the
colliding wind region being much smaller in volume and located much
closer to the B star. Hence the amount of non-thermal emission arising
in the shock cone is expected to be much smaller. The reduced
mass-loss rate from the B star and smaller volume of the resulting
shock cone explains why WR86 is found to be a weak non-thermal
emitter, while WR146 and WR147 are known strong non-thermal emitters.

\item{In none of the systems did we observe any trace of
diffuse emission down to the instrumental limit. If there is any
diffuse emission in the optical associated with the colliding wind
interface, it must be weaker than 5 10$^{-15}$ ergs cm$^{-2}$ s$^{-1}$
\AA$^{-1}$ arcsec$^{-1}$.}

\end{enumerate}

Overall, we feel that the classification of OB stars, especially the
determination of luminosity classes, is a difficult and non-trivial
task. The main reason for this is the lack of availability of a
uniform sequence of digital spectra spanning the whole spectral range
from blue to red. Existing atlases, while useful, are sometimes
fragmentary, and most are based on photographic spectra. Publication
of a comprehensive atlas of digital spectra for OB stars based on CCD
observations and covering the whole optical range from $\sim$4000 -
7000 \AA\, would be very beneficial to this field.

\acknowledgments

\clearpage

\begin{deluxetable}{llr}
\tablewidth{18pc}
\tablecaption{Line identifications in the spectrum of the WR component of WR86}
\tablehead{
\colhead{ID} & \colhead{$\lambda_{lab}$(\AA)} & \colhead{W$_{\lambda}$(\AA)}}

\startdata
 HeII          & 4338.7 & -25.0$\pm$5.0  \nl
 CIV           & 4441.5 & -31.9$\pm$0.5  \nl
 HeII          & 4541.6 & -15.0$\pm$5.0  \nl
 OIII/OV       & 5592.2/5597.9 & -31.8$\pm$2.0  \nl
 CIII          & 5695.9 & -205.$\pm$15.  \nl
 CIV           & 5801.3/5812.0 & -295.$\pm$15.  \nl
 HeI           & 5875.6 & -50.0$\pm$5.0  \nl
 HeII          & 6406.4 & -13.0$\pm$3.0  \nl
 HeII          & 6560.0 & -85.0$\pm$10.  \nl
 CIII          & 6744.4 & -120.$\pm$10.  \nl
\enddata
\end{deluxetable}

\begin{deluxetable}{lllr}
\tablewidth{22pc}
\tablecaption{Line identifications in the spectrum of the OB component of WR86}
\tablehead{
\colhead{ID} &
\colhead{$\lambda_{lab}$(\AA)} &
\colhead{$\lambda_{obs}$(\AA)} &
\colhead{W$_{\lambda}$(\AA)}
}

\startdata
H$_{\gamma}$ & 4340.5    & 4341.7 & 2.30$\pm$0.20 \nl
OII          & 4349.4    & 4349.0 & 0.70$\pm$0.20 \nl
HeI          & 4387.9    & 4389.2 & 0.55$\pm$0.05 \nl
OII          & 4414.9    & 4417.3 & 0.35$\pm$0.05 \nl
OII          & 4448.3    & 4449.3 & 0.20$\pm$0.05 \nl
HeI          & 4471.5    & 4472.4 & 1.00$\pm$0.10 \nl
MgIII        & 4479.0    & 4482.1 & 0.15$\pm$0.05 \nl
NII          & 4530.4    & 4530.7 & 0.15$\pm$0.05 \nl
SiIII        & 4553.9    & 4554.2 & 0.35$\pm$0.05 \nl
SiIII        & 4567.8    & 4569.1 & 0.30$\pm$0.05 \nl
SiIII        & 5739.7    & 5741.6 & 0.30$\pm$0.05 \nl
HeI          & 5875.6    & 5877.2 & 0.70$\pm$0.05 \nl
H$_{\alpha}$ & 6561.9    & 6564.8 & 2.00$\pm$0.05 \nl
HeI/HeII     & 6678.1/6683.2 & 6679.8 & 0.80$\pm$0.05 \nl
\enddata
\end{deluxetable}

\begin{deluxetable}{llr}
\tablewidth{18pc}
\tablecaption{Line identifications in the spectrum of the WR component of WR146}
\tablehead{
\colhead{ID} & \colhead{$\lambda_{lab}$(\AA)} & \colhead{W$_{\lambda}$(\AA)}}

\startdata
 OIII/OV       & 5592.2/5597.9 & -22.0$\pm$3.0 \nl
 CIII          & 5695.9 & -55.0$\pm$8.0 \nl
 CIV           & 5801.3/5812.0 & -600.$\pm$35. \nl
 HeI           & 5875.6 & -140.$\pm$35. \nl
 HeII          & 6406.4 & -15.0$\pm$3.0 \nl
 HeII          & 6560.0 & -165.$\pm$10. \nl
 CIII          & 6744.4 & -200.$\pm$15. \nl
\enddata
\end{deluxetable}

\begin{deluxetable}{lllr}
\tablewidth{22pc}
\tablecaption{Line identifications in the spectrum of the OB component of WR146}
\tablehead{
\colhead{ID} &
\colhead{$\lambda_{lab}$(\AA)} &
\colhead{$\lambda_{obs}$(\AA)} &
\colhead{W$_{\lambda}$(\AA)}
}

\startdata
H$_{\gamma}$  & 4340.5    & 4339.6 & 1.50$\pm$0.10 \nl
HeI           & 4387.9    & 4388.7 & 0.20$\pm$0.05 \nl
HeI           & 4471.5    & 4470.0 & 0.80$\pm$0.05 \nl
HeII          & 4541.5    & 4540.3 & 0.40$\pm$0.05 \nl
CIII          & 5695.9    & 5695.1 &-0.65$\pm$0.05 \nl
CIV           & 5811.9    & 5811.2 & 0.22$\pm$0.05 \nl
HeI           & 5875.6    & 5874.8 &  1.0$\pm$0.10 \nl
H$_{\alpha}$  & 6561.9    & 6568.3 &-4.85$\pm$0.25 \nl
HeI/HeII      & 6678.1/6683.2 & 6678.3 & 0.15$\pm$0.05 \nl
\enddata
\end{deluxetable}

\begin{deluxetable}{llr}
\tablewidth{18pc}
\tablecaption{Line identifications in the spectrum of the WR component of WR147}
\tablehead{
\colhead{ID} & \colhead{$\lambda_{lab}$(\AA)} & \colhead{W$_{\lambda}$(\AA)}}

\startdata
 NII         & 5679.6    & -6.0$\pm$1.0 \nl
 NIV         & 5736.9    & -5.8$\pm$1.0 \nl
 CIV         & 5801.3/5812.0 & -0.8$\pm$0.2 \nl
 HeI         & 5875.6    &-35.0$\pm$1.0 \nl
 HeII        & 6310.8    & -0.6$\pm$0.2 \nl
 NIV         & 6380.7    & -0.8$\pm$0.2 \nl
 HeII        & 6406.4    & -0.9$\pm$0.2 \nl
 NIII        & 6467.0/6478.7 & -3.2$\pm$0.3 \nl
 HeII        & 6527.1    & -1.0$\pm$0.1 \nl
 HeII        & 6560.0    &-45.0$\pm$1.5 \nl
 HeI/HeII    & 6678.1/6683.2 &-30.0$\pm$1.5 \nl
\enddata
\end{deluxetable}

\begin{deluxetable}{lllr}
\tablewidth{22pc}
\tablecaption{Line identifications in the spectrum of the OB component of WR147}
\tablehead{
\colhead{ID} &
\colhead{$\lambda_{lab}$(\AA)} &
\colhead{$\lambda_{obs}$(\AA)} &
\colhead{W$_{\lambda}$(\AA)}
}

\startdata
CIV           & 5801.3 & 5802.5 & 0.40$\pm$0.10 \nl
CIV           & 5811.9 & 5812.5 & 0.35$\pm$0.10 \nl
HeI           & 5875.6 & 5875.9 & 0.45$\pm$0.10 \nl
H$_{\alpha}$  & 6561.9 & 6570.8 &-4.05$\pm$0.05 \nl
\enddata
\end{deluxetable}
\begin{figure} \plotone{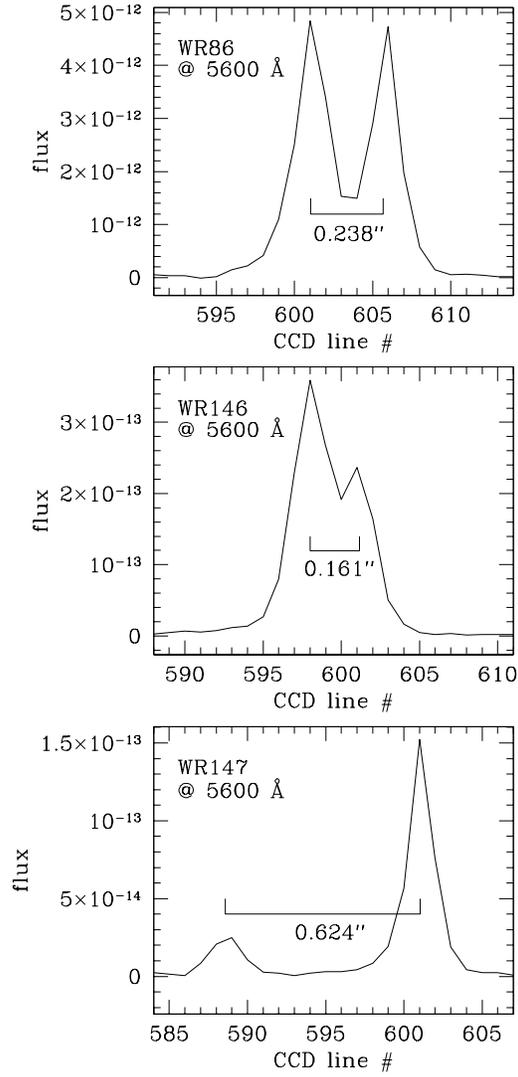}
\caption{Resolution of the WR+O systems along the slit of the STIS
camera at 5600 \AA. The slit is oriented along the line joining the
stars, with the OB star to the left and the WR star to the right on
this figure. \label{fig1}}
\end{figure}

\begin{figure} \plotone{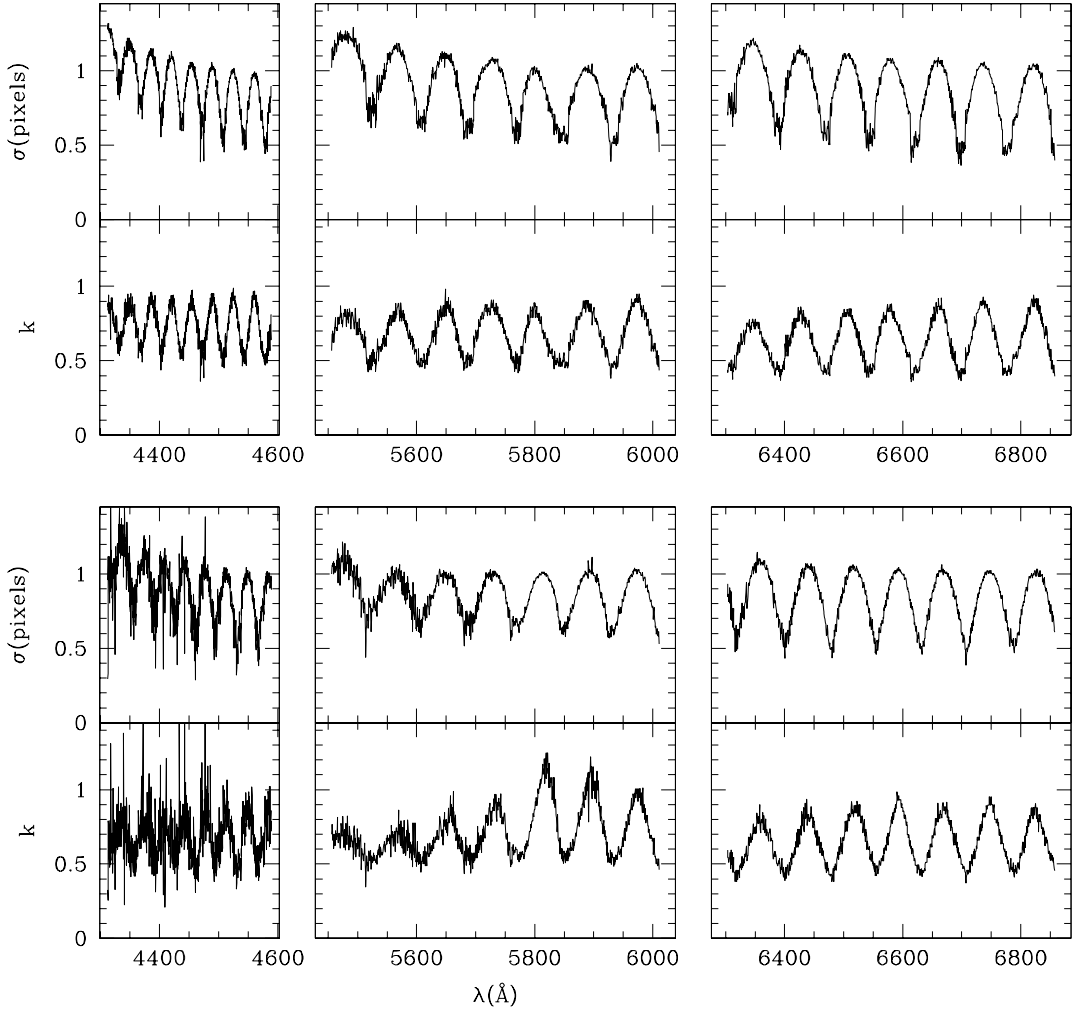}
\caption{Dispersion parameter $\sigma$ and ``peakyness'' parameter $k$
of the stellar profiles measured for the WR86 (top) and WR146 (bottom)
systems. Values are derived from a double-profile multidimensional fit
at each column on the STIS CCD image. The point spread function is
clearly dependent on the wavelength; patterns also differ
significantly between systems. The fit also yields deblended spectra
of the individual components (Figures 3-6). Deblending is reasonably
good, except at wavelengths where the WR star becomes $\gtrsim3$ times
brighter than the O star, at which point this model is not accurate
enough to yield reliable deblended spectra (see text). \label{fig2}}
\end{figure}

\begin{figure} \plotone{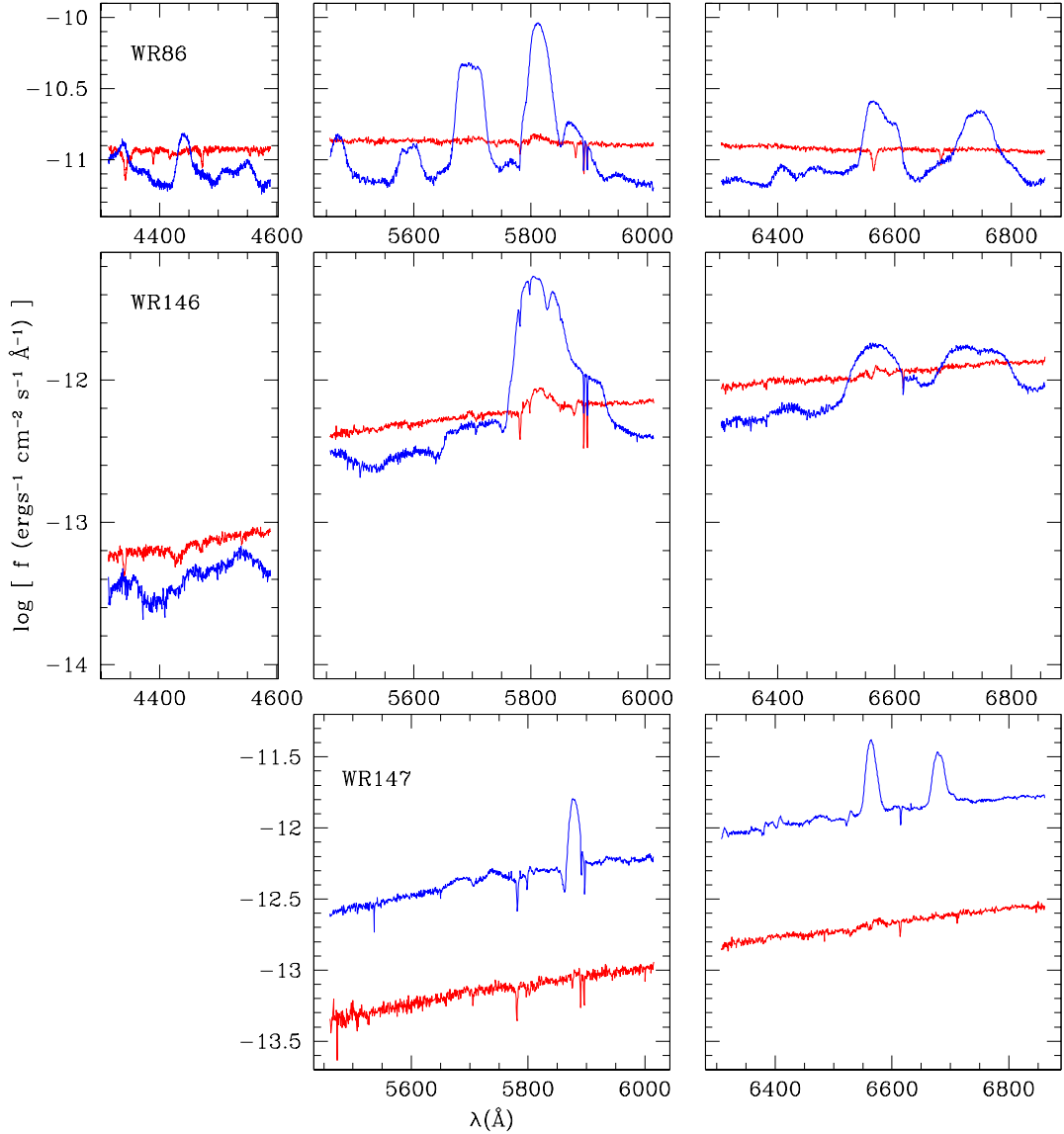}
\caption{Resolved STIS spectra of the WR86, WR146, and WR147 systems
(from top to bottom, respectively). Contamination of the OB spectrum
by the brightest of the WR star features occurred in WR86 and in WR146
(CIV $\lambda$5812 emission line) because the stars were only separated by
$\approx2$ standard deviations of the STIS point spread function. Note
the similar amounts of reddening and equivalent strengths of the
interstellar absorption features for the pairs of stars in each
system. Line identification is provided in Figures 4-9. \label{fig3}}
\end{figure}

\begin{figure} \plotone{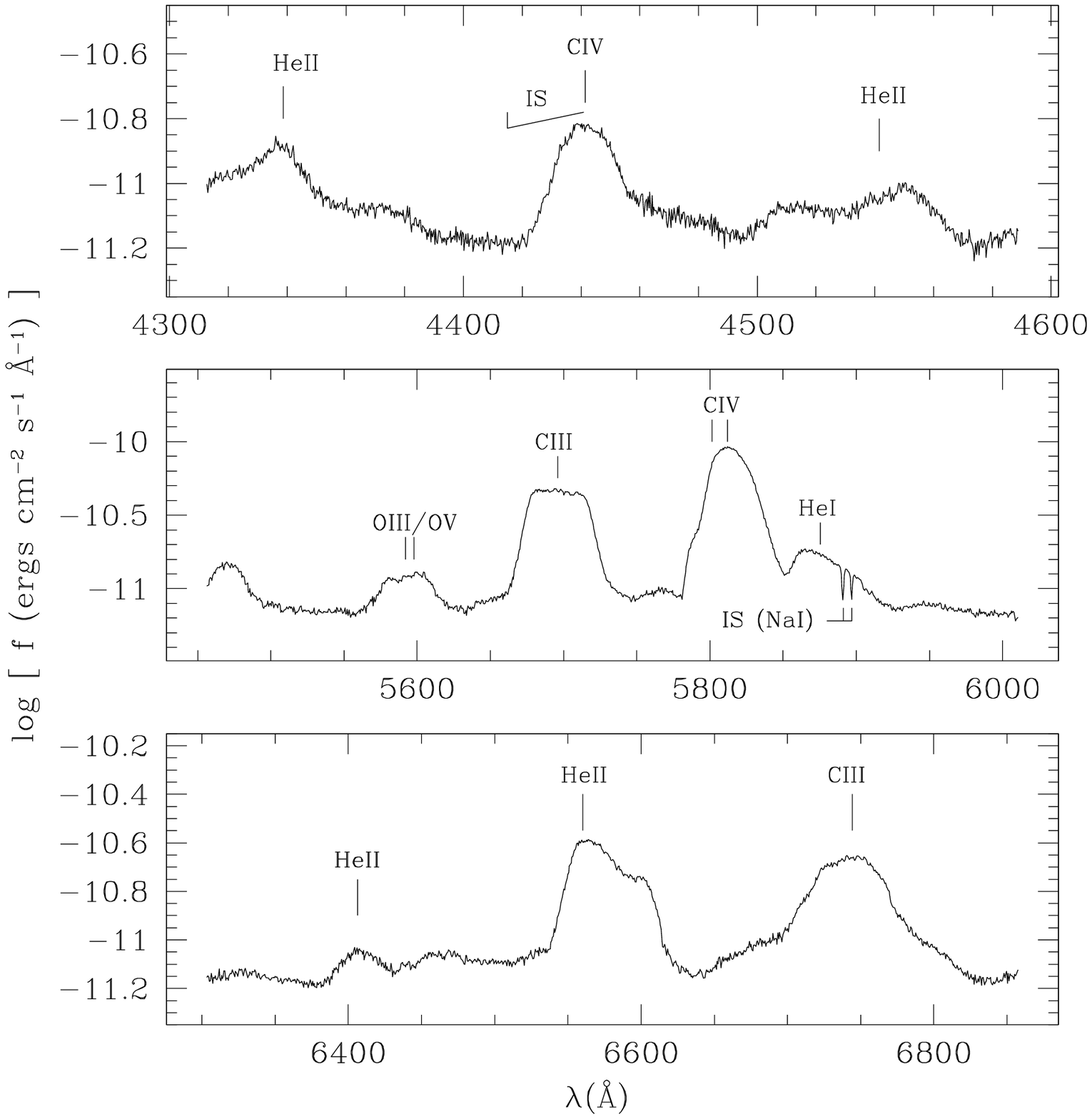}
\caption{Line identification for the WR star component in the WR86
system. Because emission-lines are very broad, most of the features
identified here are actually blends of several different lines. The
identification is for the main contributor to each feature. Equivalent
widths are listed in Table 1.\label{fig4}}
\end{figure}

\begin{figure} \plotone{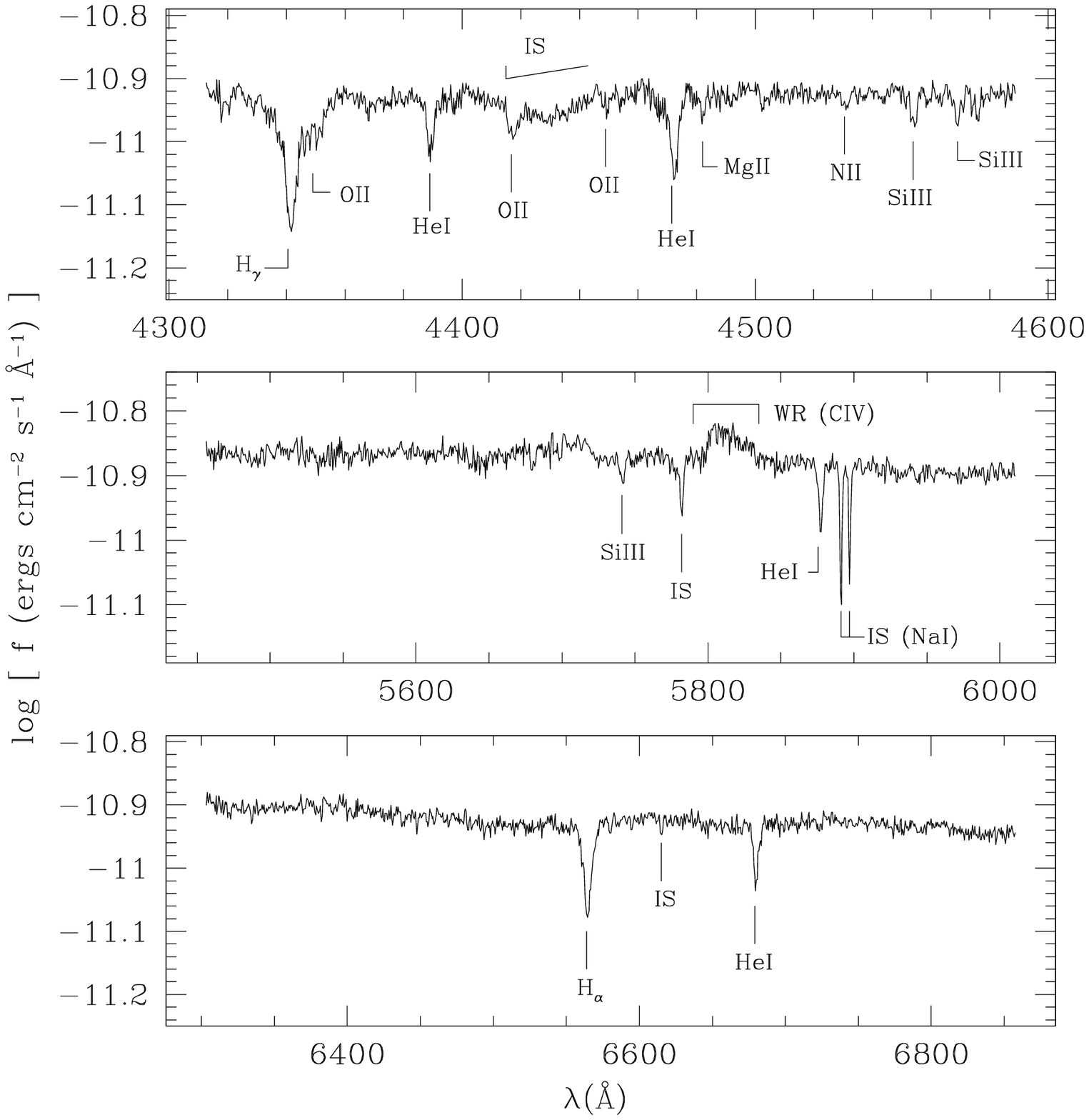}
\caption{Line identification for the B star component in the WR86
system. Contamination from the WR star is noted, along with the principal
interstellar absorption features. Equivalent widths are listed in
Table 2. The feature noted WR(CIV) is an instrumental contamination of
the light from the WR component.\label{fig5}}
\end{figure}

\begin{figure} \plotone{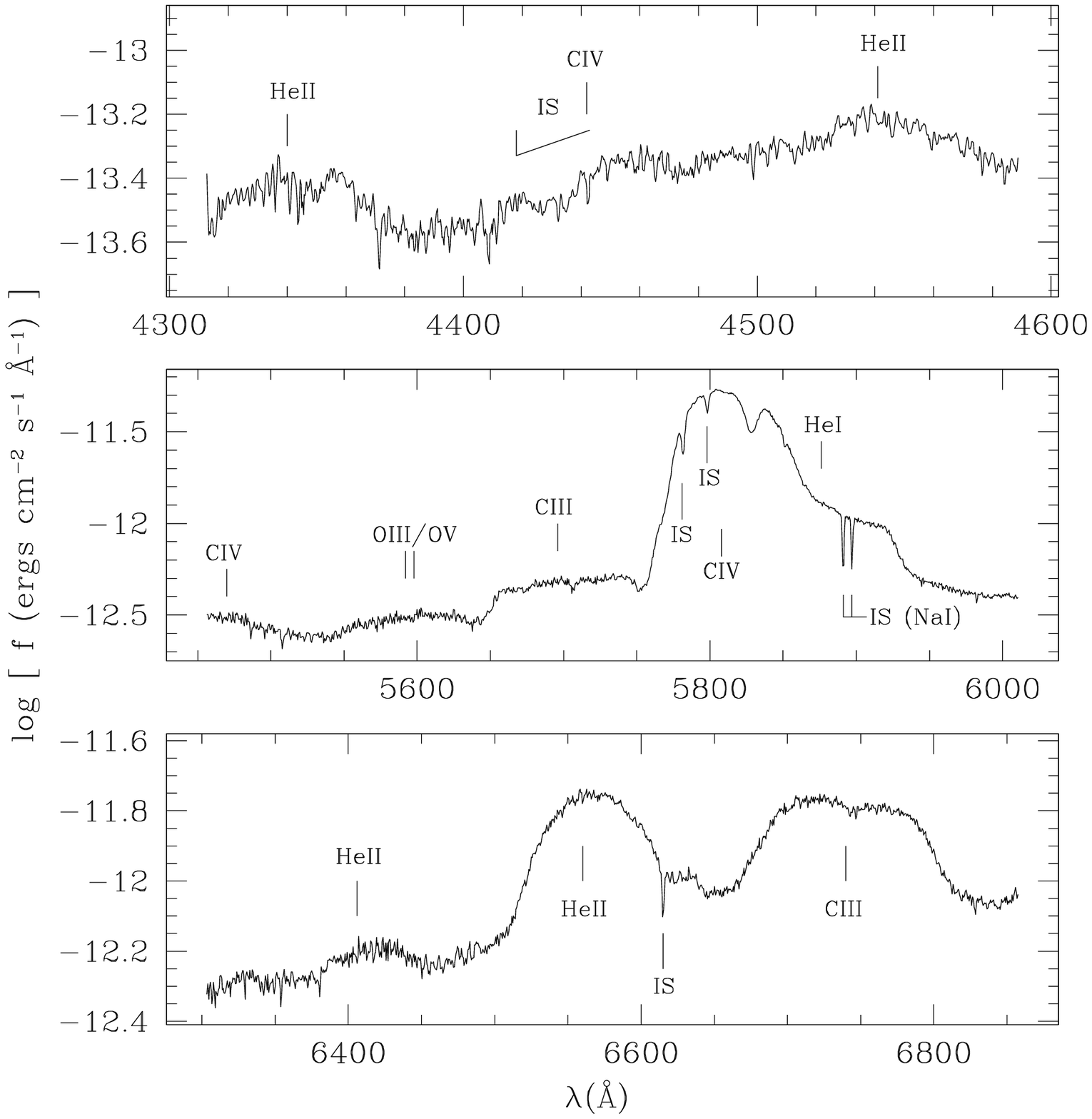}
\caption{Line identification for the WR star component in the WR146
system. Because emission-lines are very broad, most of the features
identified here are actually blends of several different lines. The
identification is for the main contributor to each feature. Equivalent
widths are listed in Table 3.\label{fig6}}
\end{figure}

\begin{figure} \plotone{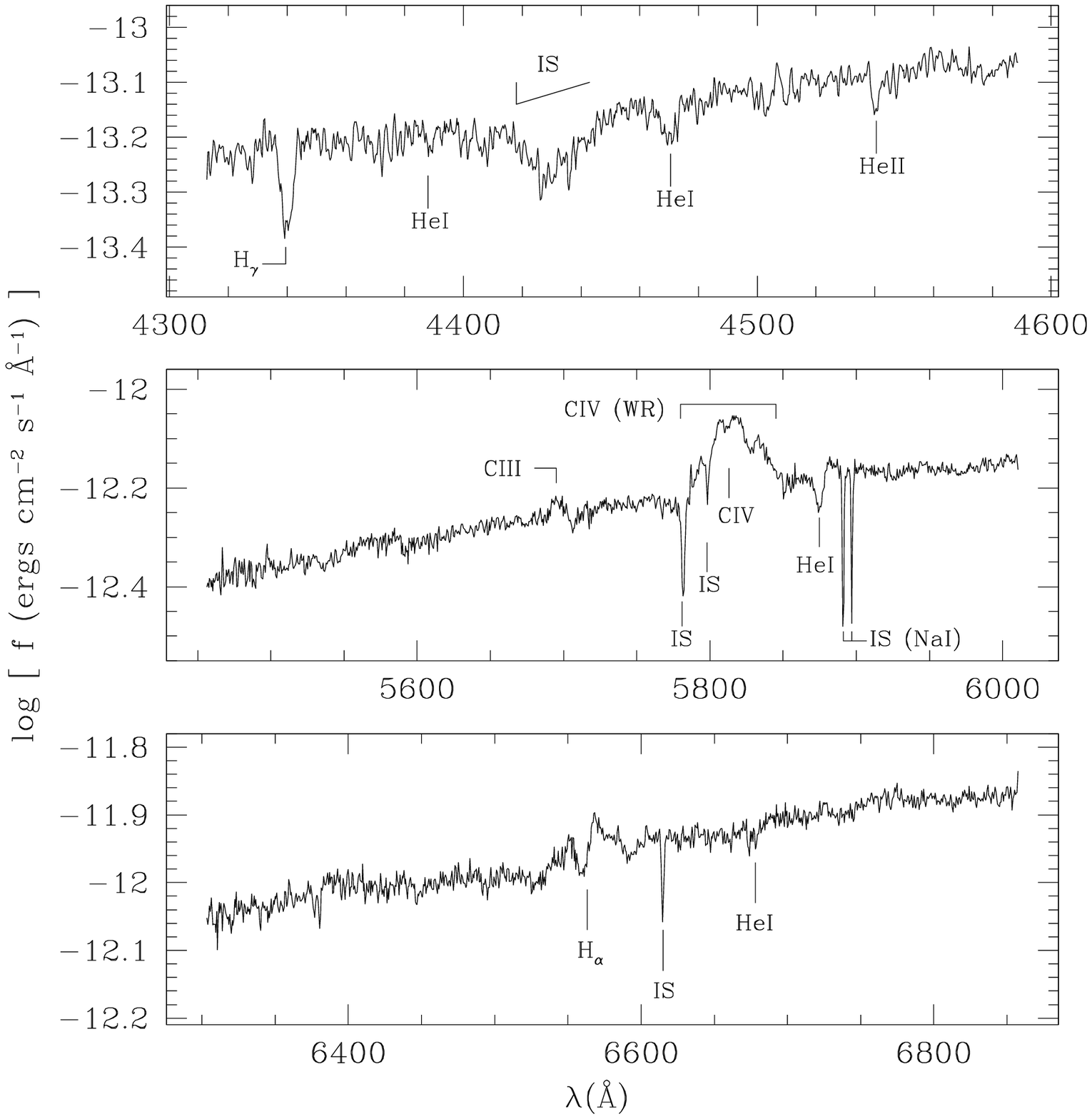}
\caption{Line identification for the O star component in the WR146
system. Contamination from the WR star is noted, along with the principal
interstellar absorption features. Equivalent widths are listed in
Table 4. The feature noted CIV(WR) is an instrumental contamination of
the light from the WR component.\label{fig7}}
\end{figure}

\begin{figure} \plotone{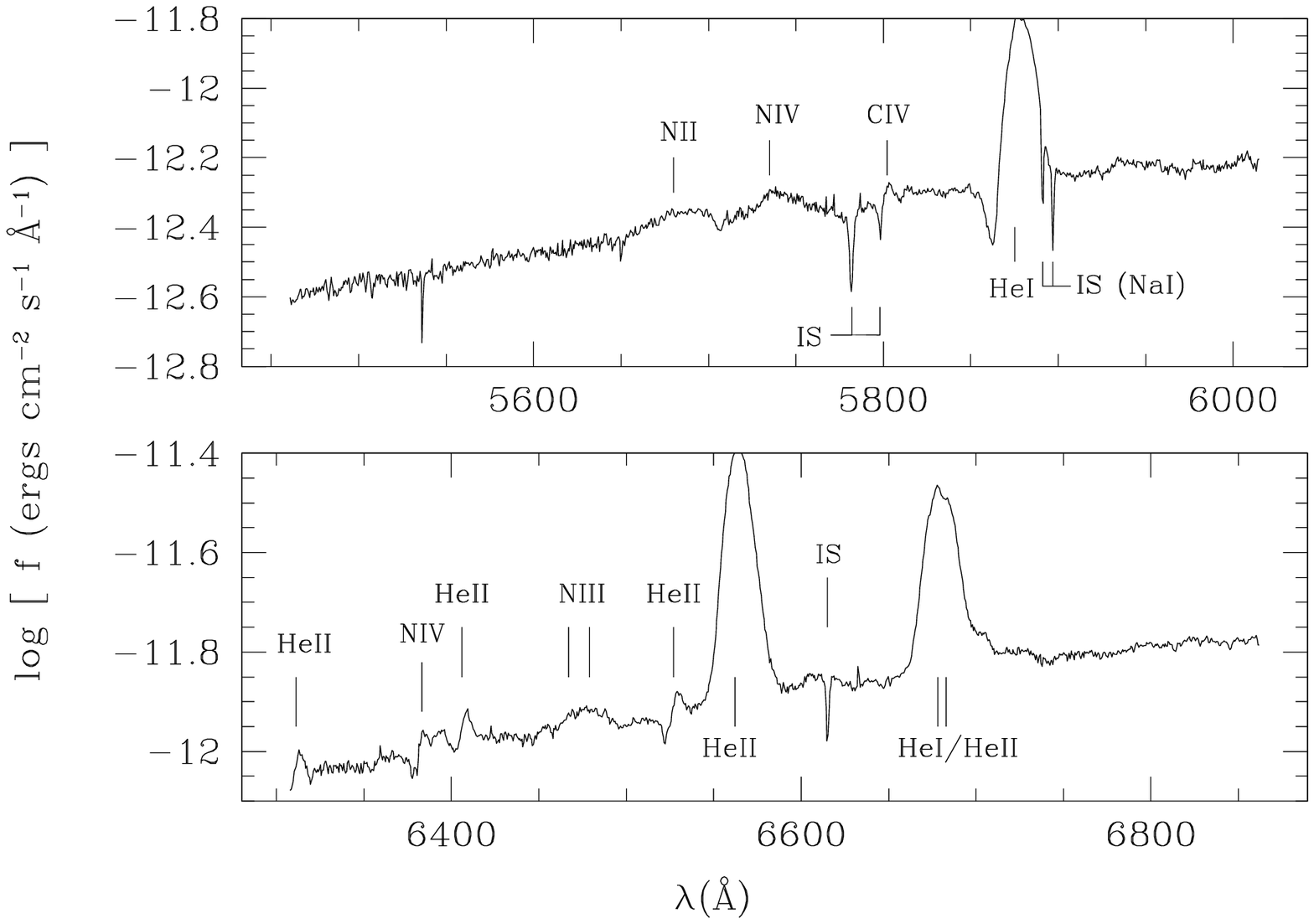}
\caption{Line identification for the WR star component in the WR147
system. Because emission-lines are very broad, most of the features
identified here are actually blends of several different lines. The
identification is for the main contributor to each feature. Equivalent
widths are listed in Table 5. \label{fig8}}
\end{figure}

\begin{figure} \plotone{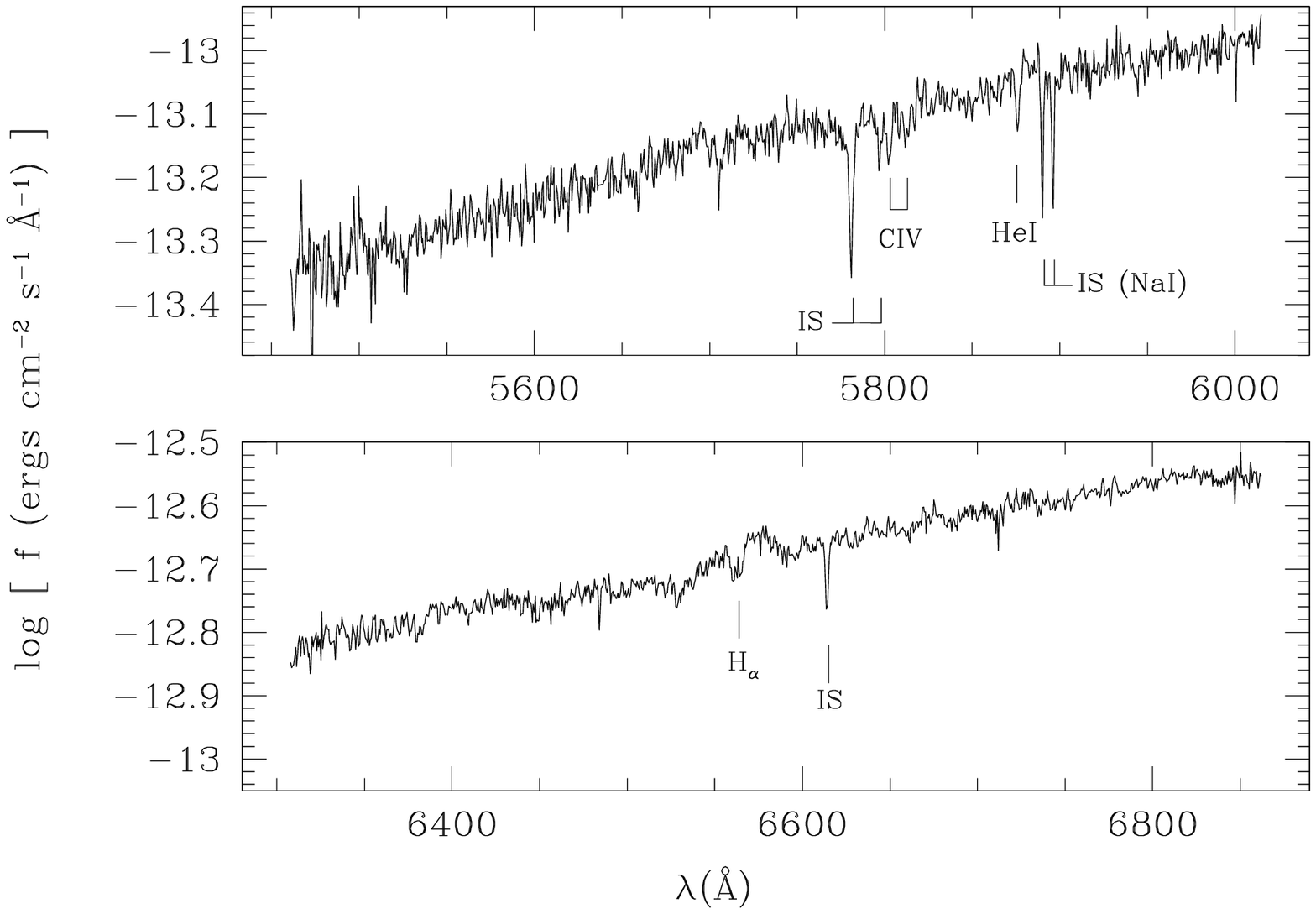}
\caption{Line identification for the O star component in the WR147
system. Well-defined intrinsic and interstellar absorption features
are noted. Equivalent widths are listed in Table 6.\label{fig9}}
\end{figure}


\begin{thebibliography}{}

\bibitem[...]{..00} Churchwell, E., Bieging, J. H., van der Hucht,
K. A., Williams, P. M., Spoelstra, T. A. Th., \& Abbott, D. C. 1992,
\apj, 393, 329

\bibitem[...]{..00} Conti, P. S., \& Alschuler, W. R. 1971, \apj, 170,
325

\bibitem[...]{..00} Conti, P. S. 1973, \apj, 179, 181

\bibitem[...]{..00} Conti, P. S. 1976, Soci\'et\'e Royale des Sciences
de Li\`ege, M\'emoires, vol. 9, 1976, p. 193-212

\bibitem[...]{..00} Crowther, P., De Marco, O., Barlow, M. J. 1998,
\mnras, 296, 367

\bibitem[...]{..00} Dougherty, S. M., Williams, P. M., van der Hucht,
K. A., Bode, M. F., \& Davis, R. J. 1996, \mnras, 280, 963

\bibitem[...]{..00} Dougherty, S. M. \& Williams, P. M. 2000, \mnras,
319, 1005

\bibitem[...]{..00} Dougherty, S. M., Williams, P. M., \& Pollacco,
D. L. 2000, \mnras, 316, 143

\bibitem[...]{..00} Eenens, P. R. J., \& Williams, P. M. 1994, \mnras,
269, 1082

\bibitem[...]{..00} Eichler, D., \& Usov, V. 1993, \apj, 402, 271

\bibitem[...]{..00} Hartkopf, W. I., Mason, B. D., Barry, D. J.,
McAlister, H. A., Bagnuolo, W. G., \& Prietro, C. M. 1993, \aj, 106,
352-360

\bibitem[...]{..00} Herrero, A., Puls, J., \& Villamariz, M. R. 2000,
\aap, 354, 193

\bibitem[...]{..00} Jeffers, H. M., van de Bos, W. H., \& Greeby,
F. M.  1963, {\it Index Catalog of Visual Double Stars, 1961.0},
Publ. Lick Obs. 21, part 1

\bibitem[...]{..00} Lundstrom, I., \& Stenholm, B. 1984, \aaps, 58, 163


\bibitem[...]{..00} Milward, C. G., \& Walker, G. A. H. 1985, \apjs,
57, 63

\bibitem[...]{..00} Massey, P., Conti, P. S., Niemela, V. S. 1981,
ApJ, 246, 145

\bibitem[...]{..00} Moffat, A. F. J. 1969, \aap, 3, 455

\bibitem[...]{..00} Moffat, A. F. J. 1995, IAU Symp. 163, Properties
of Wolf-Rayet Binaries: The Key to Understanding Wolf-Rayet Stars
(Dordrecht: Kluwer), 213

\bibitem[...]{..00} Moffat, A. F. J., Lamontagne, R., Shara, M. M.,
McAlister, H. A. 1986, \aj, 91, 1392-1399

\bibitem[...]{..00} Niemela, V. S., Shara, M. M., Wallace,
D. J., Zurek, D. R., Moffat, A. F. J. 1998, \aj, 115, 2047

\bibitem[...]{..00} Prinja, R. K., Barlow, M. J., \& Howarth,
I. D. 1990, ApJ, 361, 607

\bibitem[...]{..00} Roberts, M. S., 1962, \aj, 67, 79-85

\bibitem[...]{..00} Runacres, M. C., \& Blomme, R. 1996, \aap, 309,
544

\bibitem[...]{..00} Schmidt-Kaler, T.  1982, Numerical Data 
Functional Relationships in Science \& Technology, Landoldt-Bornstein,
New Series, Group 6, Vol. 2b, ed. K. Schaiffers \& H.H. Voigt, (Berlin:
Springer), p.1

\bibitem[...]{..00} Setia Gunawan, D. Y. A., van der Hucht, K. A., de
Bruyn, A. G., \& Williams, P. M. 2000, \aap, 356, 676

\bibitem[...]{..00} Smith, L. F. 1968, \mnras, 138, 109-121

\bibitem[...]{..00} Smith, L. F., Shara, M. M., Moffat,
A. F. J. 1990, \apj, 358, 229

\bibitem[...]{..00} Smith, L. F., Shara, M. M., Moffat, A. F. J. 1996,
\mnras, 281, 163

\bibitem[...]{..00} Turner, D. G. 1982, In: Wolf-Rayet stars:
Observations, physics, evolution; Proceedings of the Symposium,
Cozumel, Mexico, September 18-22, 1981. (A82-48127 24-90) Dordrecht,
D. Reidel Publishing Co., p. 57-60

\bibitem[...]{..00} Tuthill, P. G., Monnier, J. D., Danchi,
W. C. \nat, 398, 487

\bibitem[...]{..00} van der Hucht 1992, Astronomy and Astrophysics
Review, vol. 4, no. 2, p. 123

\bibitem[...]{..00} van der Hucht, K. A. 2001, New Astronomy Reviews, 45, 135

\bibitem[...]{..00} van der Hucht, K. A., Conti, P. S., Lundstrom, I.,
\& Stenholm, B. 1981, \ssr, 28, 227

\bibitem[...]{..00} Walborn, N. R. 1980, ApJS, 44 535

\bibitem[...]{..00} Walborn, N. R., Fitzpatrick, E. L. 1990,
\pasp, 102, 379

\bibitem[...]{..00} Wallace, D. J., Gies, D. R., Nelan, E., \&
Leitherer, C. 2000, \baas, Volume 32, No. 4, \#97.04

\bibitem[...]{..00} Wegner, W. 2000, \mnras, 319, 771

\bibitem[...]{..00} Westerlund, B. E. 1966, \apj, 145, 724

\bibitem[...]{..00} Williams, P. M., Radio emission from the stars and
the sun. ASP Conference Series, Volume 93; Proceedings of a conference
held at the University of Barcelona; Barcelona; Spain; 3-7 July 1995,
edited by A. R. Taylor and J. M. Paredes, p.15

\bibitem[...]{..00} Williams, P. M., Dougherty, S. M., Davis, R. J.,
van der Hucht, K. A., Bode, M. F., Setia Gunawan, D. Y. 1997, \mnras,
289, 10

\bibitem[...]{..00} Williams, P. M., 1999, ``Proceedings of the 193rd
symposium of the International Astronomical Union held in Puerto
Vallarta, Mexico, 3-7 November 1998". Edited by K. A. van der
Hucht, G. Koenigsberger, and P. R. J. Eenens. San Francisco,
Calif. : Astronomical Society of the Pacific, 1999., p.267

\end{thebibliography}
\end{document}